\documentclass[11pt]{article} 
\usepackage{amsmath}
\usepackage{amsfonts,amssymb}

\begin{document}

\newcommand{\nl}{\nonumber\\}
\newcommand{\nnl}{\nl[6mm]}
\newcommand{\nle}{\nl[-2.5mm]\\[-2.5mm]}
\newcommand{\nlb}[1]{\nl[-2.0mm]\label{#1}\\[-2.0mm]}
\newcommand{\ab}{\allowbreak}

\renewcommand{\leq}{\leqslant}
\renewcommand{\geq}{\geqslant}

\renewcommand{\theequation}{\thesection.\arabic{equation}}
\let\ssection=\section
\renewcommand{\section}{\setcounter{equation}{0}\ssection}

\newcommand{\be}{\bes}
\newcommand{\ee}{\ees}
\newcommand{\bes}{\begin{eqnarray}}
\newcommand{\ees}{\end{eqnarray}}
\newcommand{\eens}{\nonumber\end{eqnarray}}
\newcommand{\barr}{\begin{array}}
\newcommand{\earr}{\end{array}}

\renewcommand{\/}{\over}
\renewcommand{\d}{\partial}
\newcommand{\dAlam}{\Delta}
\newcommand{\Dslash}{\hbox{$D\kern-2.4mm/\,$}}
\newcommand{\dd}[1]{\ab\delta/\delta {#1}}
\newcommand{\ddt}{{d\/dt}}

\newcommand{\no}[1]{{\,:\kern-0.7mm #1\kern-1.2mm:\,}}

\newcommand{\mm}{{\mathbf m}}
\newcommand{\nn}{{\mathbf n}}

\newcommand{\qsmu}{q^*_\mu}
\newcommand{\qsnu}{q^*_\nu}
\newcommand{\psmu}{p_*^\mu}

\newcommand{\half}{{1\/2}}
\newcommand{\bra}[1]{\big{\langle}#1\big{|}}
\newcommand{\ket}[1]{\big{|}#1\big{\rangle}}

\newcommand{\da}{\d_\alpha}
\newcommand{\db}{\d_\beta}
\newcommand{\Ca}{C_\alpha}
\newcommand{\Cb}{C_\beta}
\newcommand{\Mab}{M^{\alpha\beta}}
\newcommand{\Kab}{K_{\alpha\beta}}

\newcommand{\fa}{\phi^\alpha}
\newcommand{\fb}{\phi^\beta}
\newcommand{\pa}{\pi_\alpha}
\newcommand{\pb}{\pi_\beta}
\newcommand{\Ea}{\EE_\alpha}
\newcommand{\Eb}{\EE_\beta}

\newcommand{\fs}{\phi^*}
\newcommand{\ps}{\pi_*}
\newcommand{\fsa}{\phi^*_\alpha}
\newcommand{\fsb}{\phi^*_\beta}
\newcommand{\psa}{\pi_*^\alpha}
\newcommand{\psb}{\pi_*^\beta}
\newcommand{\fsi}{\phi^*_i}

\newcommand{\wf}{{\overline \phi}}
\renewcommand{\wp}{{\overline \pi}}
\newcommand{\wfs}{{\overline \phi}{}^*}
\newcommand{\wps}{{\overline \pi}_*}

\newcommand{\fm}{\phi_{\mm}}
\newcommand{\fn}{\phi_{\nn}}
\newcommand{\pim}{\pi^{\mm}}
\newcommand{\pin}{\pi^{\nn}}
\newcommand{\fsm}{\phi^*_{\mm}}
\newcommand{\fsn}{\phi^*_{\nn}}
\newcommand{\psm}{\pi_*^{\mm}}
\newcommand{\psn}{\pi_*^{\nn}}

\renewcommand{\fam}{\phi^\alpha_\mm}
\newcommand{\fbn}{\phi^\beta_\nn}
\newcommand{\pam}{\pi_\alpha^\mm}
\newcommand{\wfam}{{\overline \phi}{}^\alpha_\mm}
\newcommand{\wfbn}{{\overline \phi}{}^\beta_\nn}
\newcommand{\wpam}{{\overline \pi}{}_\alpha^\mm}

\newcommand{\wfm}{{\overline \phi}_{\mm}}
\newcommand{\wfn}{{\overline \phi}_{\nn}}
\newcommand{\wpm}{{\overline \pi}^{\mm}}
\newcommand{\wpn}{{\overline \pi}^{\nn}}
\newcommand{\wfsm}{{\overline \phi}^*_{\mm}}
\newcommand{\wfsn}{{\overline \phi}^*_{\nn}}
\newcommand{\wpsm}{{\overline \pi}_*^{\mm}}
\newcommand{\wpsn}{{\overline \pi}_*^{\nn}}

\newcommand{\wfa}{{\overline \phi}{}^\alpha}
\newcommand{\wpa}{{\overline \pi}_\alpha}
\newcommand{\wpb}{{\overline \pi}_\beta}
\newcommand{\wfsa}{{\overline \phi}^*_\alpha}
\newcommand{\wpsa}{{\overline \pi}_*^\alpha}
\newcommand{\wpsb}{{\overline \pi}_*^\beta}
\newcommand{\wfsam}{{\overline \phi}^*_{\alpha,\mm}}
\newcommand{\wfsan}{{\overline \phi}^*_{\alpha,\nn}}
\newcommand{\wfsbn}{{\overline \phi}^*_{\beta,\nn}}
\newcommand{\wpsam}{{\overline \pi}_*^{\alpha,\mm}}
\newcommand{\wpsan}{{\overline \pi}_*^{\alpha,\nn}}
\newcommand{\wpsbm}{{\overline \pi}_*^{\beta,\mm}}
\newcommand{\wpsbn}{{\overline \pi}_*^{\beta,\n}}

\newcommand{\fsam}{\phi^*_{\alpha,\mm}}
\newcommand{\fsbn}{\phi^*_{\beta,\nn}}
\newcommand{\psam}{\pi_*^{\alpha,\mm}}
\newcommand{\psbn}{\pi_*^{\beta,\nn}}
\newcommand{\Eam}{\EE_{\alpha,\mm}}

\newcommand{\si}{\sigma}
\newcommand{\eps}{\epsilon}
\newcommand{\dlt}{\delta}
\newcommand{\om}{\omega}
\newcommand{\al}{\alpha}
\newcommand{\bt}{\beta}
\newcommand{\gm}{\gamma}
\newcommand{\ka}{\kappa}
\newcommand{\la}{\lambda}
\newcommand{\vth}{\vartheta}
\renewcommand{\th}{\theta}
\newcommand{\rep}{\varrho}

\newcommand{\vect}{{\mathfrak{vect}}}
\newcommand{\map}{{\mathfrak{map}}}

\newcommand{\im}{{\rm im}\ }
\newcommand{\e}{{\rm e}}
\renewcommand{\div}{{\rm div}}
\newcommand{\afn}{{\rm afn\,}}
\newcommand{\til}{{\tilde{\ }}}

\newcommand{\eikx}{\e^{ik\cdot x}}
\newcommand{\dNx}{{d^N \kern-0.2mm x\ }}
\newcommand{\dNxp}{{d^N \kern-0.2mm x'\ }}
\newcommand{\dNxb}{{d^N \kern-0.2mm x''\ }}
\newcommand{\dNk}{{d^N \kern-0.2mm k\ }}
\newcommand{\dmu}{{\d_\mu}}

\newcommand{\larroww}[1]{{\ \stackrel{#1}{\longleftarrow}\ }}
\newcommand{\summ}[1]{\sum_{|\mm|\leq #1}}
\newcommand{\intdm}{\int_{-\infty}^\infty dm\ }

\newcommand{\Np}[1]{{N+p\choose N #1}}
\newcommand{\Npr}{{N+p-r\choose N-r}}

\newcommand{\tr}[1]{{\rm tr}_{#1}\kern0.7mm}
\newcommand{\oj}{{\mathfrak g}}
\newcommand{\hh}{{\mathfrak h}}

\newcommand{\J}{{\cal J}}
\newcommand{\U}{{\cal U}}
\newcommand{\N}{{\cal N}}
\newcommand{\QQ}{{\cal Q}}
\newcommand{\PP}{{\cal P}}
\newcommand{\EE}{{\cal E}}
\newcommand{\FF}{{\cal F}}
\newcommand{\HH}{{\cal H}}
\newcommand{\II}{{\cal I}}
\newcommand{\GG}{{\cal G}}

\newcommand{\cl}{{cl}}
\newcommand{\qm}{{qm}}
                                            
\newcommand{\TT}{{\mathbb T}}
\newcommand{\RR}{{\mathbb R}}
\newcommand{\CC}{{\mathbb C}}
\newcommand{\ZZ}{{\mathbb Z}}
\newcommand{\NN}{{\mathbb N}}

\title{{Manifestly covariant canonical quantization I: the free scalar
field}}

\author{T. A. Larsson \\
Vanadisv\"agen 29, S-113 23 Stockholm, Sweden\\
email: thomas.larsson@hdd.se}

\maketitle 
\begin{abstract} 
Classical physics is reformulated as a constrained Hamiltonian system in
the history phase space. Dynamics, i.e. the Euler-Lagrange equations,
play the role of first-class constraints. This allows us to apply
standard methods from the theory of constrained Hamiltonian systems, e.g.
Dirac brackets and cohomological methods. In analogy with BRST
quantization, we quantize in the history phase space first and 
impose dynamics afterwards. To obtain a truly covariant
formulation, all fields must be expanded in a Taylor series around the
observer's trajectory, which acquires the status of a quantized physical
field. The formalism is applied to the harmonic oscillator and to the free
scalar field. Standard results are recovered, but only in the
approximation that the observer's trajectory is treated as a classical
curve.
\end{abstract}

\vskip 3 cm
PACS (2003): 03.65.Ca, 03.70.+k, 11.10.Ef.


\newpage

\section{Introduction} 

The path-integral formulation of quantum mechanics has one major 
advantage over the Hamiltonian formalism, namely that it preserves
manifest covariance. However, the canonical formalism may justly be
considered more fundamental. It would therefore be desirable to have a
Hamiltonian formulation of quantum mechanics which maintains manifest
covariance. This paper initiates a program towards such a formulation.

The first observation is the well-known fact that phase space is really a
covariant concept; it is the space of solutions to the classical
equations of motion, i.e. the space of histories. A phase space point
$(q,p)$ corresponds to the history $(q(t),p(t))$, such that $(q(0),p(0))
= (q,p)$. Since $p(t)$ is completely determined from $q(t)$ and
Hamilton's equations, one usually drops reference to $p(t)$ and define
the covariant phase space $\Sigma$ to be the space of histories 
$q(t)$ that solve the Euler-Lagrange (EL) equations.

We are not interested so much in $\Sigma$ itself, but rather in
the space of functions over $\Sigma$, $C(\Sigma)$, which will become our
Hilbert space after quantization. This space can be identified with the
space of functions over history space $\QQ$, which is spanned by
arbitrary trajectories $q(t)$, modulo the ideal
$\N$ generated by the EL equations: $C(\Sigma) = C(\QQ)/\N$.
It is a standard result in the antifield formalism that $C(\Sigma)$ can
described as a resolution of a certain differential complex \cite{HT92}. 
In the absense of gauge symmetries, this complex is known as the
Koszul-Tate (KT) complex.

Alas, the antifield formalism is not suited for canonical quantization, 
although this is of course no problem if we only want to do path
integrals. After adding antifields, the history space $\QQ$ is replaced
by an extended history space $\QQ^*$. We can define an antibracket in
$\QQ^*$, but in order to do canonical quantization we need an honest
Poisson bracket. To this end, we introduce canonical momenta conjugate to
the history and its antifield, and obtain an even larger space $\PP^*$,
which may be thought of as the phase space corresponding to the 
extended history space $\QQ^*$.
It turns out to be possible to define a KT complex also in $C(\PP^*)$.
The KT differential can now be written as a bracket, $\dlt F = [Q,F]$.
Under a technical assumption, all momenta are killed in cohomology, and
we obtain a different description of $C(\Sigma)$.

Since $C(\PP^*)$ is naturally equipped with a Poisson bracket, we can now
do canonical quantization on $\PP^*$ by replacing Poisson brackets with
commutators. However, we also need to represent the graded Heisenberg
algebra on a Hilbert space, in a way which makes the Hamiltonian bounded
from below. At this step we must single out a privileged time direction
and give up manifest covariance. The Hamiltonian is simply defined as the
generator of rigid time translations. Since it commutes with
the KT operator $Q$ even after normal ordering, the Hamiltonian is well
defined in cohomology. This process defines a non-covariant quantization
in the covariant phase space.

Covariant quantization requires a covariant definition of the Hamiltonian.
To this end, we introduce {\em the observer's trajectory} in spacetime,
$ q^\mu(t)$. All fields and antifields are expanded in a Taylor series
around $ q^\mu(t)$, before canonical momenta are introduced and 
the KT complex is constructed. The parameter $t$
has no physical significance, so the generator of rigid $t$ translations
can not serve as a Hamiltonian; indeed, the fields in $\Sigma$ are
$t$-independent. Instead the Hamiltonian is the operator which translates
the fields relative to the observer. This definition turns out to
agree with the usual Hamiltonian in the limit that $ q^\mu(t)$ can be
regarded as a classical variable.

Hence the key idea is to regard the physical phase space as the
constraint surface in the history phase space $\PP$, with the
EL equations playing the role of a first-class constraint.
One way to treat a constrained system is to introduce a gauge-fixing
condition, which relates momenta to velocities, replace Poisson by
Dirac brackets, and eliminate the constraints (dyna\-mics) prior to 
quantization. This strategy is investigated in Section \ref{sec:Dirac}.

In the following two sections, we describe how to construct a
cohomological model for the covariant space space, and how to do
non-covariant and covariant canonical quantization, respectively, before
imposing dyna\-mics. In the next three sections, the formalism is applied
to the harmonic oscillator and the free scalar field, in non-covariant
and covariant form, respectively. The result deviates from conventional
canonical quantization in two respects, one technical and one 
substantial.

1. The momenta are only eliminated in cohomology provided that the
Hessian, i.e. the second functional derivative matrix of the action, is
non-singular. This technical assumption is not true for the harmonic
oscillator. I argue that one can avoid this problem, which is present in
the standard antifield treatment as well, by adding a small perturbation
to the action, making the Hessian non-singular. In the limit that the
perturbation vanishes, the non-covariant quantization yields the correct
Hilbert space. Without this trick, the Hilbert space describes quanta of
the right energy, but there would be too many types of quanta.

2. The energy of a plane wave is the projection of its momentum $k_\mu$
along the observer's velocity, i.e. $\dot q^\mu(t) k_\mu$. In general, this
is a quantum operator which creates an observer quantum from the vacuum.
This is a c-number (and the correct c-number) only if the observer can be
regarded as a classical object, i.e. if the reference state is a mixed
state with many observer quanta. But this is only an approximation. 
In a fundamental theory, everything must be quantized, including the
observer's trajectory. Classical observation is merely a special limit
of quantum interaction. It is therefore encouraging that treating
canonical quantization in a manifestly covariant way forces us to 
introduce a quantized observer trajectory.

In section \ref{sec:repar} the algebra of reparametrizations of the
observer's trajectory is discussed. This plays no role if we explicitly
break reparametrization freedom by requiring the observer to move along a
straight line in Minkowski space. In applications to general-covariant
theories the observer will move along a geodesic instead, keeping
reparametrization invariance intact. The reparametrization algebra will
acquire quantum corrections, making it into a Virasoro algebra. A
well-defined central extension can only be constructed in three
dimensions and with twice as many fermionic as bosonic fields.

The last section contains a discussion of the results and an outlook
for future work.

\section{Phase space as a constraint surface in history phase space}
\label{sec:Dirac}

Consider a classical dynamical system with action $S$ and degrees of
freedom $\fa$. As is customary in the antifield literature, we use an
abbreviated notation where the index $\alpha$ stands for both discrete
indices and spacetime coordinates.
Dynamics is governed by the Euler-Lagrange (EL) equations,
\be
\Ea = \da S \equiv {\dlt S\/\dlt\fa} = 0.
\label{EL}
\ee
An important role is also played by the Hessian, i.e. the symmetric 
second functional-derivative matrix
\be
\Kab = K_{\bt\al} = \db\Ea \equiv {\dlt \Ea\/\dlt\fb} 
= {\dlt^2 S\/\dlt\fa\dlt\fb}.
\label{Hess}
\ee
The Hessian is assumed non-singular, so it has an
inverse $\Mab$ satisfying
\be
K_{\bt\gm}M^{\gm\al} = M^{\al\gm}K_{\gm\bt} = \dlt^\al_\bt.
\ee

In the Hamiltonian formulation we split the index $\al = (i, t)$ and
consider the phase space $\Sigma$ with coordinates $(\phi^i, \pi_j)$.
$\Sigma$ is equipped with a Poisson bracket satisfying the commutation
relations
\bes
[\pi_j, \phi^i] &=& \dlt^i_j, 
\nlb{Heis}
{[}\phi^i, \phi^j] &=& [\pi_i, \pi_j]  = 0.
\eens
The time evolution of the system is described by a curve 
$(\phi^i(t), \pi_j(t)) \in \Sigma$, governed by Hamilton's equations,
\be
{d\phi^i(t)\/dt} = [\phi^i(t), H], \qquad
{d\pi_j(t)\/dt} = [\pi_j(t), H].
\ee
The Hamiltonian formalism breaks manifest covariance, due to the special
role played by the time coordinate. However, the phase space is a
covariant concept; it is the space of solutions to the classical equations
of motions $(\phi^i(t), \pi_j(t)) = (\fa,\pa)$. The standard way to
coordinatize such a curve is to use the initial values $(\phi^i, \pi_j) =
(\phi^i(0), \pi_j(0))$, but physics does not depend on this particular way
to put coordinates on $\Sigma$. Since $\pa$ is completely determined from
$\fa$ and Hamilton's equations, we may drop reference to $\pa$ and define
the covariant phase space $\Sigma$ to be the space of histories $\fa$
modulo the the EL equations.

Let $\QQ$ be the space of histories $\fa$. For each history $\fa$ we
introduce the canonical momentum $\pa = \dd{\fa}$, and define $\PP$ to be
the space of histories $(\fa,\pb)$. $\PP$ has a natural symplectic
structure, given by the  canonical commutation relations
\be
[\pb,\fa] = \dlt^\al_\bt, \qquad
[\fa,\fb] = [\pa,\pb] = 0.
\ee
Spelled out in detail, it reads
\bes
[\pi_j(t), \phi^i(t')] &=& \dlt^i_j \dlt(t-t'), 
\nle
{[}\phi^i(t), \phi^j(t')] &=& [\pi_i(t), \pi_j(t')]  = 0.
\eens
Observe that this Heisenberg algebra holds in the history phase space,
not in the physical phase space to be constructed. In particular, the
factor $\dlt(t-t')$ means that the Poisson brackets are not only 
defined at equal times.

This suggests that we may regard the EL equation $\Ea = 0$ as a constraint
in $\PP$, and use standard methods for constrained Hamiltonian systems to
recover the physical phase space $\Sigma$. However, some care must be
exercised. The EL equations do not determine a historu uniquely, but
only up to initial conditions. We therefore redefine 
$\fa \to (\fa,\phi^i)$, where $\phi^i$ are the initial conditions
(typically, $\phi(0)$ and $\dot\phi(0)$), and $\fa$ are the remaining
variables ($\phi(t)$ for $t>0$).

We thus impose dyna\-mics as a constraint $\Ea\approx0$ in $\PP$. As is
customary, $\approx$ denotes weak equality, i.e. equality when all
constraints have been taken into account. This constraint is first class;
$[\Ea,\Eb] = 0 \approx 0$ since $\Ea$ only depends on the $\phi$'s but
not on the $\pi$'s. To make the constraint second class, we must
introduce more constraints which make the Poisson bracket matrix
non-singular. We take $\pa \approx \Ca$, where $\Ca(\phi)$ is some
function which is independent of $\pa$. The precise choice is not
important; it will affect the definition of canonical momenta but
dyna\-mics for the fields is always given by the same EL equation.
There is no field equations for the initial conditions, but their
canonical momenta are constrained by $\pi_i \approx C_i$.

A set of constraints $\chi_A\approx0$ are second class if
the Poisson bracket matrix $\Delta_{AB}=[\chi_A,\chi_B]$ is invertible;
denote the inverse by $\Delta^{AB}$.
The Dirac bracket
\be
[F,G]_* = [F,G] - [F,\chi_A] \Delta^{AB} [\chi_B,G]
\label{Dirac}
\ee
defines a new Lie bracket which is compatible with the 
constraints: $[F, \chi_A]_* = 0$ for every $F\in C(\PP)$. Since this
is a strong equality, i.e. it holds throughout $C(\PP)$ and not only
modulo constraints, we can now solve the constraints $\chi_A=0$.
The Dirac bracket becomes the Poisson bracket on the reduced
phase space.

The second-class constraints in the history phase space are
\bes
\Ea &\approx& 0, \nl
\pa - \Ca &\approx& 0, \\
\pi_i - C_i &\approx& 0. 
\eens
The Poisson-bracket matrix reads
\bes
\Delta_{AB}
&=& \Big[\begin{pmatrix} \Ea \\ {\pa-\Ca} \\ {\pi_i-C_i} \end{pmatrix}, 
\begin{pmatrix} \Eb & {\pb-\Cb} & {\pi_j-C_j} \end{pmatrix}\Big] \nle
&=& \begin{pmatrix} 
0 & -\db\Ea & -\d_j\Ea \\ 
\da\Eb & \db\Ca-\da\Cb & \d_j\Ca-\da C_j \\
\d_i\Eb & \db C_i-\d_i\Cb & \d_j C_i-\d_i C_j
\end{pmatrix}.
\eens
Assuming that $\d_i\Eb = \d_i\Cb = 0$, it has the inverse
\be
\Delta^{AB} =
\begin{pmatrix} 
\Gamma^{\al\bt} & \Mab & 0 \\
-\Mab & 0 & 0 \\
0 & 0 & \Omega^{ij}
\end{pmatrix},
\ee
where $\Mab$ is the inverse of the Hessian,  
$\Omega^{ij}$ is the inverse of $\Omega_{ij} = \d_j C_i-\d_i C_j$, and 
$\Gamma^{\al\bt} = M^{\al\gm}(\d_\dlt C_\gm - \d_\gm C_\dlt) M^{\dlt\bt}
= M^{\al\gm}\Omega_{\gm\dlt} M^{\dlt\bt}$.

One easily verifies that the Dirac brackets commute with the constraints,
\be
[F, \Ea]_* = [F, \pa-\Ca]_* = [F, \pi_i-C_i]_* = 0,
\ee
for every $F\in C(\PP)$. We can thus solve the constraints for $\fa$,
$\pa$ and $\pi_i$. The constraint surface $\Sigma$ is identified with
the physical phase space. It has coordinates $\phi^i$ and the Poisson
bracket is given by
\be
[\phi^i, \phi^j]_* = \Omega^{ij}.
\ee
Alternatively, we can solve the constraint $\pi_i-C_i\approx0$ for 
$\phi^i$, and describe $\Sigma$ in terms of coordinates $\pi_i$ with
Poisson bracket
\be
{[}\pi_i, \pi_j]_* = \d_i C_k \Omega^{k\ell} \d_j C_\ell.
\ee
The Hamiltonian in $\PP$ is simply the generator of rigid time
translations along the trajectories. When we pass to Dirac brackets,
in acquires knowledge about the dyna\-mics of the system.

This section illustrates how the physical phase space $\Sigma$ can be 
obtained from the history phase space $\PP$:
\begin{enumerate}
\item
Introduce the Euler-Lagrange equations as first class constraints.
\item
Introduce gauge conditions which relate momenta to velocities, making
the constraints second class.
\item
Pass to Dirac bracket and eliminate the constraints.
\end{enumerate}
We will not pursue this strategy further, because relating momenta to
velocities necessarily breaks covariance. In the next section
we introduce the analogue of the BRST formalism, where $\Sigma$ is
constructed by cohomological methods, and no gauge-fixing condition is
necessary at any stage.

\section{Koszul-Tate cohomology}
\label{sec:KT}

In the previous section we described dyna\-mics as a constraint in the
history phase space, and recovered the physical phase space by solving
this constraint. A more elegant way to treat constrained systems is the
BRST approach. Here one never solves any constraint. Instead, a nilpotent
BRST operator is introduced in a extended phase space including ghosts,
and the physical phase space is identified with the zeroth cohomology
group of the BRST complex. We saw in the previous section that the EL
equations may be viewed as a first-class constraint in the history phase
space. Although we assume that there are no gauge symmetries, the
situation is nevertheless very similar here, with the EL equation playing
the role of a gauge symmetry. There should thus be an analogue of the
BRST formalism in the history phase space. In particular, we want to
construct a differential complex in $C(\PP)$, which is a resolution of
the space of functions over the true phase
space, $C(\Sigma) = C(\QQ)/\N$.

Let us recall how this is done in the antifield formalism \cite{HT92}.
Introduce an antifield $\fsa$ for each EL equation $\Ea=0$, and
replace the space of $\phi$-histories $\QQ$ by the extended history
space $\QQ^*$, spanned by both $\phi$ and $\fs$.
In $\QQ^*$ we define the Koszul-Tate (KT) differential $\dlt$ by
\be
\dlt \fa = 0, \qquad \dlt \fsa = \Ea.
\ee
One checks that $\dlt$ is nilpotent, $\dlt^2 = 0$.
Let the antifield number $\afn \fa = 0$, $\afn \fsa = 1$. The KT
differential clearly has antifield number $\afn\dlt = -1$.

The space $C(\QQ^*)$ decomposes into subspaces $C^k(\QQ^*)$ of
fixed antifield number
\be
C(\QQ^*) = \sum_{k=0}^\infty C^k(\QQ^*)
\ee
The KT complex is 
\be
0 \larroww \dlt C^0 \larroww \dlt C^1 \larroww \dlt 
C^2 \larroww \dlt \ldots
\label{complex1}
\ee
The cohomology spaces are defined as usual by
$H_\cl^\bullet(\dlt) = \ker\dlt/\im\dlt$, i.e.
$H_\cl^k(\dlt) = (\ker\dlt)_k/(\im\dlt)_k$, where
the subscript $\cl$ indicates that we deal with a classical phase space.
It is easy to see that
\bes
(\ker \dlt)_0 &=& C(\QQ), \nle
(\im \dlt)_0 &=& C(\QQ)\Ea \equiv \N.
\eens
Thus $H_\cl^0(\dlt) = C(\QQ)/\N = C(\Sigma)$.
Since we assume that there are no non-trivial relations among the
$\Ea$, the higher cohomology groups vanish. This is a standard
result \cite{HT92}. The complex (\ref{complex1}) thus gives us
a resolution of the covariant phase space $C(\Sigma)$, which by
definition means that
$H_\cl^0(\dlt) = C(\Sigma)$, $H_\cl^k(\dlt) = 0$, for all $k>0$.

The extended history space $\QQ^*$ is not a phase space, because it has
no Poisson bracket. It does admit an antibracket, defined on the
coordinates by
\be
(\fa,\fsb) = \dlt^\al_\bt.
\label{antibracket}
\ee
However, the antibracket is not an proper Poisson bracket because it is
symmetric, $(F,G) = +(G,F)$, so is can not be used for canonical
quantization. Moreover, as we will see in the next section, the
antibracket can not even be defined for jets.

The key observation is now that the same space $C(\Sigma)$ admits a
different resolution.
Introduce canonical momenta $\pa = \dd{\fa}$ and 
$\pa = \dd{\fa}$ for both the fields and antifields. The momenta
satisfy by definition the graded canonical commutation relations
\bes
[\pb,\fa] = \dlt^\al_\bt, &\qquad&
[\fa,\fb] = [\pa,\pb] = 0, 
\nlb{ccr*}
[\psb,\fsa]_+ = \dlt_\al^\bt, &\qquad&
[\fsa,\fsb]_+ = [\psa,\psb]_+ = 0,
\eens
where $[\cdot,\cdot]_+$ is the symmetric bracket.
Let $\PP$ be the phase space of histories with basis $(\fa,\pb)$,
and let $\PP^*$ be the extended phase space with basis
$(\fa,\pb,\fsa,\psb)$. 
The following table summarize the different spaces that we have defined.
\[
\barr{|c|l|l|}                                                       
\hline
\hbox{Space} & \hbox{Basis} & \hbox{Name}\\
\hline
\Sigma & \fa: \Ea = 0 & \hbox{Physical phase space}\\
\QQ & \fa  & \hbox{History space}\\
\QQ^* & \fa, \fsa & \hbox{Extended history space} \\
\PP & \fa, \pb & \hbox{History phase space} \\
\PP^* & \fa, \pb, \fsa, \psb & \hbox{Extended history phase space} \\
\hline
\earr
\]

The definition of the KT differential extends to $\PP^*$ by 
requiring that
$\dlt F = [Q,F]$ for every $F \in C(\PP^*)$, where the KT operator is
\be
Q = \Ea \psa.
\ee
It acts on the various fields as
\bes
\dlt \fa &=& 0, \nl
\dlt \fsa &=& \Ea, 
\nlb{dltfp}
\dlt \pa &=& -{\dlt \Eb\/\dlt\fa}\psb = - \Kab \psb, \nl
\dlt \psa &=& 0,
\eens
where $\Kab$ is the Hessian (\ref{Hess}).
We check that $\dlt$ is still nilpotent: $\dlt^2 = [Q,Q]_+ = 0$.

Like $C(\QQ)$, the function space $C(\PP^*)$
decomposes into subspaces of fixed antifield number,
$C(\PP^*) = \sum_{k=-\infty}^\infty C^k(\PP^*)$.
We can therefore define a KT complex in $C(\PP^*)$
\be
\ldots \larroww Q C^{-2} \larroww Q C^{-1} \larroww Q 
C^0 \larroww Q C^1 \larroww Q 
C^2 \larroww Q \ldots
\label{complex2}
\ee
Because the Hessian (\ref{Hess}) is non-singular by assumption
with inverse $\Mab$, we can invert the relation 
$\dlt \pa = -\Kab\psb$ 
and get
\be
\psa = -\Mab\dlt\pb = \dlt(-\Mab\pb),
\ee
since $\Mab$ depends on $\phi$ alone.

Let us now compute the cohomology.
Any function which contains $\pa$ is not closed, so
$\ker\dlt = C(\phi,\fs,\ps)$.
Moreover, $\im\dlt$ is generated by the two ideals
$\Ea$ and $\psa$. The momenta $\pa$ and $\psa$ thus vanish
in cohomology, and the part with zero antifield number is thus still
$H_\cl^0(\dlt) = C(\QQ)/\N = C(\Sigma)$.
The higher cohomology groups $H_\cl^k(\dlt) = 0$ by the same argument
as above. Hence the complex (\ref{complex2}) yields a different 
resolution of the same phase space $\Sigma$. 

It is important that the spaces $C^k$ in (\ref{complex2}) are phase
spaces, equipped with the Poisson bracket (\ref{ccr*}). Unlike the
resolution (\ref{complex1}), the new resolution (\ref{complex2}) therefore
allows us to do canonical quantization: replace Poisson brackets by
commutators and represent the graded Heisenberg algebra (\ref{ccr*}) on 
a Hilbert space. However, the Heisenberg algebra can be represented on
different Hilbert spaces. To pick the correct one, we must impose the
physical condition that there is an energy which is bounded on below.

To define the
Hamiltonian, we must single a privileged variable $t$ among the $\al$'s,
and declare it to be time.
Thus replace $\al = (i,t)$, so e.g. $\fa = \phi^i(t)$,
$\Ea = \EE_i(t)$, etc.
This step means of course that we sacrifice covariance.
The Hamiltonian reads
\be
H = -i\int dt\ \dot\phi^i(t)\pi_i(t) + \dot\fsi(t)\ps^i(t).
\label{Hamc}
\ee
It satisfies
\bes
[H, \phi^i(t)] = -i\dot\phi^i(t), &\qquad&
[H, \pi_i(t)] = -i\dot\pi_i(t), \nle
{[}H, \fsi(t)] = -i\dot\fsi(t), &\qquad&
[H, \ps^i(t)] = -i\dot\ps^i(t).
\eens

Expand all fields in a Fourier series with respect to time, e.g,
\be
\phi^i(t) = \int_{m=-\infty}^\infty dm\ \phi^i(m) \e^{imt}.
\ee
The Fourier modes $\pi_i(m)$, $\fsi(m)$ and $\ps^i(m)$ are defined
analogously.
The Hamiltonian acts on the Fourier modes as
\bes
[H, \phi^i(m)] = m\phi^i(m), &\qquad&
[H, \pi_i(m)] = m\pi_i(m), \nle
{[}H, \fsi(m)] = m\fsi(m), &\qquad&
[H, \ps^i(m)] = m\ps^i(m).
\eens

Now quantize. In the spirit of BRST quantization, our strategy is to
quantize first and impose dyna\-mics afterwards.
In the extended history phase space $\PP^*$, we
define a Fock vacuum $\ket 0$ which is annihilated by all negative
frequency modes, i.e.
\be
\phi^i(-m)\ket 0 = \pi_i(-m)\ket 0 = \fsi(-m)\ket 0 
= \ps^i(-m)\ket 0 = 0,
\ee
for all $-m < 0$. We must also decide which of the zero modes that
annihilate the vacuum, but the decision is not important unless
zero-momentum modes will survive in cohomology, and even then it will
not affect the eigenvalues of the Hamiltonian.

The Hamiltonian (\ref{Hamc}) does not act in a well-defined manner, 
because it assigns an infinite energy to the Fock vacuum. To correct
for that, we replace the Hamiltonian by
\be
H = -i\int dt\ \no{\dot\phi^i(t)\pi_i(t)} + \no{\dot\fsi(t)\ps^i(t)},
\label{Hamq}
\ee
where normal ordering $\no{\cdot}$ moves negative frequency modes to
the right and positive frequency modes to the left. The vacuum has 
zero energy as measured by the normal-ordered Hamiltonian, $H\ket0 = 0$.
The Hilbert space can be identified with
\be
\HH(\PP^*) = C(\phi^i(m>0), \pi_i(m>0), \fsi(m>0), \ps^i(m>0)).
\ee
The energy of a state in $\HH(\PP^*)$ follows from
\be
H \phi^{i_1}(m_1)...\ps^{i_n}(m_n)\ket 0 
= (m_1 + ... + m_n)  \phi^{i_1}(m_1)...\ps^{i_n}(m_n)\ket 0.
\label{Hnoncov}
\ee

It is important that the KT operator
\be
Q = \Ea\psa = \int dt\ \EE_i(t) \ps^i(t) = \intdm \EE_i(m)\ps^i(-m)
\ee
is already normal ordered, because $\Ea$ and $\psa$ commute. 
This means that $Q^2 = 0$ also quantum mechanically; there are no
anomalies. Moreover, $Q$ still commutes with the Hamiltonian,
$[Q, H] = 0$, and this property is not destroyed by normal 
ordering.
Hence the Hilbert space $\HH(\PP^*)$ has also a well-defined 
decomposition into subspaces of definite antifield number,
\be
\HH(\PP^*) = ... + \HH^{-2} + \HH^{-1} + \HH^0 + \HH^1 + \HH^2 + ...
\ee
There is a KT complex in $\HH(\PP^*)$
\be
\ldots \larroww Q \HH^{-2} \larroww Q \HH^{-1} \larroww Q 
\HH^0 \larroww Q \HH^1 \larroww Q 
\HH^2 \larroww Q \ldots
\label{complexq}
\ee
The physical Hilbert space is identified with
$\HH(\Sigma) = \HH_\qm^0(Q) = (\ker Q)_0/(\im Q)_0$.
The action of the Hamiltonian on the physical Hilbert space is still 
given by (\ref{Hnoncov}), restricted to $\HH(\Sigma)\subset\HH(\PP^*)$,
and that coincides with the conventional action of the Hamiltonian.

Hence we have quantized the theory given by the EL equation (\ref{EL})
by first quantizing the space of phase space histories $\PP^*$, and
then imposing dyna\-mics through KT cohomology.

\section{Covariant phase space, covariant quantization}
\label{sec:covar}

A covariant definition of the phase space was given in the previous
section, but the Hamiltonian and thus the quantum Hilbert space broke
covariance, due to the selection of the privileged time coordinate.
In this section we correct this defect.

The compact notation in Section \ref{sec:KT} is not very useful
here, because the notion of covariance does not make sense unless
some indices are identified with spacetime coordinates. So we assume
that we have some fields $\fa(x)$, where $x = (x^\mu) \in \RR^N$ is
the spacetime coordinate. The EL equations read
\be
\Ea(x) \equiv {\dlt S\/\dlt \fa(x)} = 0.
\ee
We also need the Hessian
\be
\Kab(x,x') = K_{\bt\al}(x',x) 
= {\dlt \Ea(x)\/\dlt\fb(x')} 
= {\dlt^2 S\/\dlt\fa(x)\dlt\fb(x')}.
\label{Hess2}
\ee
which we assume is non-singular. 

Now let all fields depend on an additional parameter $t$. It will
eventually be identified with time, but so far it is completely 
unrelated to the $x^\mu$.
Upon the substitution $\fa(x) \to \fa(x,t)$,
the EL equations are replaced by
\be
\Ea(x,t) = 0.
\label{EL2}
\ee
The Hessian (\ref{Hess2}) becomes
\be
\Kab(x,t,x',t') = K_{\bt\al}(x',t',x,t) 
= {\dlt \Ea(x,t)\/\dlt\fb(x',t')},
\ee
which has the inverse
$\Mab(x,t,x',t')$ satisfying
\bes
&&\int \dNxb \int dt'' K_{\bt\gm}(x,t,x'',t'')M^{\gm\al}(x'',t'',x',t') \nl
&=& \int \dNxb \int dt'' M^{\al\gm}(x,t,x'',t'')K_{\gm\bt}(x'',t'',x',t') 
\\
&=& \dlt^\al_\bt \dlt(x-x') \dlt(t-t').
\eens

To remove the condition (\ref{EL2}) in cohomology we introduce
antifields $\fsa(x,t)$.
But the fields in the physical phase space do not depend on the parameter
$t$, which gives rise to the extra condition
\be
\d_t\fa(x,t) \equiv {\d\fa(x,t)\/\d t} = 0.
\ee
We can implement this condition by introducing new antifields $\wfa(x,t)$.
However, the identities $\d_t\Ea(x,t) \equiv 0$ give rise to unwanted
cohomology. To kill this condition, we must introduce yet another
antifield $\wfsa(x,t)$.
The KT differential $\dlt$ is defined by
\bes
\dlt \fa(x,t) &=& 0, \nl
\dlt \fsa(x,t) &=& \Ea(x,t), \nle
\dlt \wfa(x,t) &=& \d_t\fa(x,t), \nl
\dlt \wfsa(x,t) &=& \d_t\fsa(x,t)
- \int\dNxp \int dt' \Kab(x,t,x',t') \fsb(x',t').
\eens
The zeroth cohomology group $H_\cl^0(\dlt)$ equals $C(\phi)$, modulo
the ideals generated by $\Ea(x,t)$ and $\d_t\fa(x,t)$.
Moreover, the wouldbe cohomology related to the identity
\be
\dlt \Big(\d_t\fsa(x,t) - 
\int\dNxp \int dt' {\dlt \Ea(x,t)\/\dlt\fb(x',t')} \fsb(x',t') \Big) 
\equiv 0
\ee
is killed because the RHS equals $\dlt\wfsa(x,t)$.

Introduce canonical momenta for all fields and antifields:
$\pa(x,t) = \dd{\fa(x,t)}$, $\psa(x,t) = \dd{\fsa(x,t)}$,
$\wpa(x,t) = \dd{\wfa(x,t)}$, and $\wpsa(x,t) = \dd{\wfsa(x,t)}$,
with non-zero commutation relations
\bes
[\pb(x,t),\fa(x',t')] &=& \dlt^\al_\bt \dlt(x-x') \dlt(t-t'), \nl
{[}\psb(x,t),\fsa(x',t')]_+ &=& \dlt_\al^\bt \dlt(x-x') \dlt(t-t'), \nle
{[}\wpb(x,t),\wfa(x',t')]_+ &=& \dlt_\al^\bt \dlt(x-x') \dlt(t-t'), \nl
{[}\wpsb(x,t),\wfsa(x',t')] &=& \dlt^\al_\bt \dlt(x-x') \dlt(t-t').
\eens
The KT operator takes the explicit form
\be
Q &=& \int \dNx \int dt\ \Big( \Ea(x,t))\psa(x,t) + \d_t\fa(x,t)\wpa(x,t) 
\\
&&+( \d_t\fsa(x,t)
- \int\dNxp \int dt' \Kab(x,t,x',t') \fsb(t'))\wpsa(x,t) \Big).
\eens
From this we can read off the action of $\dlt$ on the momenta.
As in the previous section, the zeroth cohomology group consists of
functions $\fa(x,t)$ which satisfy $\Ea(x,t) = 0$ and $\d_t\fa(x,t)=0$.
Hence $H_\cl^0(\dlt) = C(\Sigma)$, as desired.

At this point, we must define a Hamiltonian. The candidate
\bes
H_0 &=& -i\int dt\ \Big(\d_t\fa(x,t)\pa(x,t) + \d_t\fsa(x,t)\psa(x,t) 
\nlb{Hamconstr}
&&+ \d_t\wfa(x,t)\wpa(x,t) + \d_t\wfsa(x,t)\wpsa(x,t) \Big)
\eens
might seem natural, but it is not acceptable. The action of the
Hamiltonian is KT exact, e.g.
\be
[H_0, \fa(x,t)] = \d_t\fa(x,t) = \dlt\wfa(x,t),
\ee
and thus $H_0\approx0$. This $H_0$ is not a genuine Hamiltonian, but
rather a Hamiltonian constraint $H_0\approx0$, familiar from canonical 
quantization of general relativity.

However, we can construct a well-defined and physical Hamiltonian with
some extra work. The crucial idea is to introduce the observer's
trajectory $ q^\mu(t) \in \RR^N$, and then expand all fields in a Taylor
series around this trajectory. The Taylor coefficients and
the observer's trajectory together constitute a jet, more precisely the
infinite jet, corresponding to the field.
We can now define a genuine Hamiltonian which moves the fields relative
to the observer.
As explained in \cite{Lar98},
the same step is also crucial in the representation theory of algebras
of diffeomorphisms and gauge transformations.

Hence we make the Taylor expansion
\be
\fa(x,t) = \sum_{\mm} {1\/\mm!} \fam(t)(x-q(t))^\mm,
\label{Taylor}
\ee
where $\mm = (m_1, \ab m_2, \ab ..., \ab m_N)$, all $m_\mu\geq0$, is a 
multi-index of length $|\mm| = \sum_{\mu=1}^N m_\mu$,
$\mm! = m_1!m_2!...m_N!$, and
\be
(x-q(t))^\mm = (x^1-q^1(t))^{m_1} (x^2-q^2(t))^{m_2} ...
 (x^N-q^N(t))^{m_N}.
\label{power}
\ee
Denote by $\mu$ a unit vector in the $\mu$:th direction, so that
$\mm+\mu = (m_1, \ab ...,m_\mu+1, \ab ..., \ab m_N)$, and let
\be
\fam(t) = \d_\mm\fa(q(t),t)
= \underbrace{\d_1 .. \d_1}_{m_1} .. 
\underbrace{\d_N .. \d_N}_{m_N} \fa(q(t),t)
\label{jetdef}
\ee
be the $|\mm|$:th order derivative of $\fa(x,t)$ evaluated on the
observer's trajectory $ q^\mu(t)$. 

The Taylor coefficients $\fam(t)$ are referred to as 
{\em jets}; more precisely, infinite jets. Similarly, we
define a $p$-jet by truncation to $|\mm|\leq p$. We will not
need finite $p$-jets in this paper, but they play an important role
as a regularization of symmetry generators.
A note on nomenclature may be appropriate at this point.
A $p$-jet is usually defined as an equivalence class of functions; two 
functions are equivalent if all mixed partial derivatives at a given
point $q$, up to order $p$, agree. Since each equivalence class
has a unique representative which is a polynomial of order at most $p$,
namely the truncated Taylor series around $q$, we will identify the jet
with the Taylor series. Moreover, the function $\fam(t)$ defines a
trajectory in jet space, but for brevity we will simply refer to 
$\fam(t)$ itself as a jet.

Expand also the Euler-Lagrange equations and the antifields in
a similar Taylor series,
\bes
\Ea(x,t) &=& \sum_{\mm} {1\/\mm!} \Eam(t)(x-q(t))^\mm, \nle
\fsa(x,t) &=& \sum_{\mm} {1\/\mm!} \fsam(t)(x-q(t))^\mm,
\eens
etc. These relations define the jets $\Eam(t)$, $\fsam(t)$, $\wfam(t)$
and $\wfsam(t)$. Jets of antifields will sometimes be called antijets.

The equation of motion and the time-independence condition translate
into
\bes
\Eam(t) &=& 0, \nle
D_t\fam(t) &\equiv& \ddt\fam(t) - \sum_\mu \dot q^\mu(t)\fa_{\mm+\mu} = 0.
\eens
The KT differential $\dlt$ which implements these conditions is
\bes
\dlt \fam(t) &=& 0, \nl
\dlt \fsam(t) &=& \Eam(t), 
\nle
\dlt \wfam(t) &=& D_t\fam(t), \nl
\dlt \wfsam(t) &=& D_t\fsam(t) 
 - \sum_\nn \int dt'\ K^\nn_{\mm;\al\bt}(t,t') \wfbn(t').
\eens
The cohomology group $H_\cl^0(\dlt)$ consists of linear combinations
of jets $\fam(t)$ satisfying $\Eam(t)=0$ and $D_t\fam(t)=0$.

The Taylor expansion requires that we introduce the observer's
trajectory as a physical field, but what equation
of motion does it obeys? The obvious answer is the geodesic equation,
which we compactly write as $\GG_\mu(t) = 0$. The geodesic operator
$\GG_\mu(t)$ is a function of the metric $g_{\mu\nu}(q(t),t)$
and its derivatives on the curve $ q^\mu(t)$. To eliminate this ideal
in cohomology we introduce the trajectory antifield $\qsmu(t)$, and
extend the KT differential to it:
\bes
\dlt  q^\mu(t) &=& 0, 
\nlb{geo}
\dlt \qsmu(t) &=& \GG_\mu(t).
\eens
For models defined over Minkowski spacetime, the geodesic equation
simply becomes $\ddot q^\mu(t) = 0$, and the KT differential reads
\be
\dlt \qsmu(t) = \eta_{\mu\nu} q^\nu(t).
\ee
$H_\cl^0(\dlt)$ only contains trajectories which are straight lines,
\be
 q^\mu(t) = u^\mu t + a^\mu,
\label{line}
\ee
where $u^\mu$ and $a^\mu$ are constant vectors. We may also require
that $u^\mu$ has unit length, $u_\mu u^\mu = 1$. This condition fixes
the scale of the parameter $t$ in terms of the Minkowski metric, so
we may regard it as proper time rather than as an arbitrary parameter.

Now introduce the canonical momenta $\pam(t)$, $\psam(t)$, $\wpam(t)$,
$\wpsam(t)$ for the jets and antijets (jet and antijet momenta),
and momenta $ p_\mu(t)$ and $\psmu(t)$ for the observer's trajectory 
and its antifield. The defining relations are
\bes
[\pam(t), \fbn(t')] &=& \dlt_\al^\bt \dlt^\mm_\nn \dlt(t-t'), \nl
{[}\psam(t), \fsbn(t')] &=& \dlt^\al_\bt \dlt^\mm_\nn \dlt(t-t'), \nl
{[}\wpam(t), \wfbn(t')] &=& \dlt_\al^\bt \dlt^\mm_\nn \dlt(t-t'), \nle
{[}\wpsam(t), \wfsbn(t')] &=& \dlt^\al_\bt \dlt^\mm_\nn \dlt(t-t'), \nl
{[} p_\mu(t),  q^\nu(t')] &=& \dlt_\mu^\nu \dlt(t-t'), \nl
{[}\psmu(t), \qsnu(t')] &=& \dlt^\mu_\nu \dlt(t-t').
\eens
The advantage of this formalism is that we can now define a
genuine Hamiltonian $H$, which translates the fields relative to the
observer or vice versa. Since the formulas are shortest when $H$ acts
on the trajectory but not on the jets, we make that choice, and define
\be
H = i\int dt\ (\dot q^\mu(t) p_\mu(t) + \dot\qsmu(t)\psmu(t)).
\label{H2}
\ee
Note the sign; moving the fields forward in $t$ is equivalent to moving
the observer backwards.
This Hamiltonian acts on the jets as
\bes
&&[H,  q^\mu(t)] = i\dot q^\mu(t), \qquad
[H,  p_\mu(t)] = i\dot p_\mu(t), \nl
&&{[}H, \qsmu(t)] = i\dot\qsmu(t), \qquad
[H, \psmu(t)] = i\dot\psmu(t), \\
&&{[}H, \fam(t)] = [H, \fsam(t)] 
 = [H, \wfam(t)] = [H, \wfsam(t)] = 0, \nl
&&{[}H, \pam(t)] = [H, \psam(t)] 
 = [H, \wpam(t)] = [H, \wpsam(t)] = 0.
\eens
Substituting this formula into (\ref{Taylor}), we get the energy of
the fields from
\be
[H, \fa(x,t)] = -i\dot q^\mu(t)\dmu\fa(x,t).
\label{Hcl}
\ee
This a crucial result, because it allows us to define a
genuine energy operator in a covariant way. 
In Minkowski space, the trajectory is a straight line (\ref{line}),
and $\dot q^\mu(t) = u^\mu$. If we take $u^\mu$ to be the constant
four-vector $u^\mu = (1,0,0,0)$, then (\ref{Hcl}) reduces to
\be
[H, \fa(x,t)] = -i {\d\/\d x^0}\fa(x,t).
\ee
Equation (\ref{Hcl}) is thus a genuine covariant generalization of
the energy operator.

Is there an analogue of the antibracket (\ref{antibracket}) in jet
space? The answer is no. The obvious candidate would be
\be
(\fam(t), \fsbn(t')) = \dlt^\al_\bt \dlt_{\mm,\nn} \dlt(t-t').
\ee
However, the RHS contains an object $\dlt_{\mm,\nn}$ with two lower
multi-indices, and such an object transforms non-trivially. Hence the
antibracket in jet space is not well defined.

Now we quantize the theory.
Since all operators depend on the parameter $t$, we can define
the Fourier components, e.g. 
\be
\fam(t) = \intdm \fam(m)\e^{imt}, \qquad
 q^\mu(t) = \intdm  q^\mu(m)\e^{imt}.
\ee
The the Fock vacuum $\ket 0$ is defined to be annihilated by
all negative frequency modes, i.e.
\bes
&&\fam(-m)\ket 0 = \fsam(-m)\ket 0 
= \wfam(-m)\ket 0 = \wfsam(-m)\ket 0 = 0, \nl
&&\pim(-m)\ket 0 = \psm(-m)\ket 0 
= \wpm(-m)\ket 0 = \wpsm(-m)\ket 0 
= 0, 
\label{LER}\\
&& q^\mu(-m)\ket 0 = \qsmu(-m)\ket 0 
=  p_\mu(-m)\ket 0 = \psmu(-m)\ket 0 = 0,
\eens
for all $-m < 0$.

The normal-ordered form of the Hamiltonian (\ref{H2}) reads, in
Fourier space,
\be
H = -\intdm m( \no{ q^\mu(m) p_\mu(m)} + \no{\qsmu(m) p_\mu(-m)} ),
\label{Hq}
\ee
where double dots indicate normal ordering with respect to frequency.
This ensures that $H\ket 0 = 0$.
The classical phase space $H_\cl^0(\dlt)$ is thus the the space of
fields $\fa(x)$ which solve $\Ea(x)=0$,
and trajectories $ q^\mu(t) = u^\mu t+a^\mu$, where $u^2 = 1$.
After quantization, the fields and trajectories become operators which
act on the physical Hilbert space $\HH = H_\qm^0(Q)$, which is the space
of functions of the positive-energy modes of the classical phase
space variables.

This construction differs technically from conventional canonical
quantization, but there is also a physical difference. Consider the
state $\ket{\fa(x)} = \fa(x)\ket 0$ which excites one $\phi$ quantum
from the vacuum. The Hamiltonian yields
\bes
H \ket{\fa(x)} &=& -i\dot q^\mu(t)\dmu\fa(x)\ket 0 \nl
&=& -i\ket{\dot q^\mu(t)\dmu\fa(x)} \\
&=& -i\ket{u^\mu\dmu\fa(x)}.
\eens
If $u^\mu$ were a classical variable, the state $\ket{\fa(x)}$ would
be a superposition of energy eigenstates:
\be
H \ket{\fa(x)} = -iu^\mu\dmu\ket{\fa(x)}.
\ee
In particular, let $u^\mu = (1,0,0,0)$ be a unit vector in the $x^0$
direction and $\fa(x) = \exp(ik\cdot x)$ be a plane wave. We then 
define the state $\ket{0;u,a}$ by
\be
 q^\mu(t)\ket{0;u,a} = (u^\mu t + a^\mu)\ket{0;u,a}.
\label{ua}
\ee
Now write $\ket{k; u,a} =\exp(ik\cdot x)\ket{0;u,a}$ for 
the single-quantum energy eigenstate.
\be
H \ket{k; u,a} = k_\mu u^\mu\ket{k;u,a},
\ee
so the eigenvalue of the Hamiltonian is $k_\mu u^\mu = k_0$, as
expected. Moreover, the lowest-energy condition (\ref{LER}) ensures
that only quanta with positive energy will be excited; if 
$k_\mu u^\mu < 0$ then $\ket{k; u,a} = 0$.

However, the present analysis shows that it is in principle wrong to
consider $u^\mu$ and $a^\mu$ as classical variables. The definition
(\ref{ua}) means that the reference state $\ket{0;u,a}$ is a very
complicated, mixed, macroscopic state where the observer moves along a
well-defined, classical trajectory. This is of course an excellent
approximation in practice, but it is in principle wrong.

\section{Harmonic oscillator}

The action and Euler-Lagrange equations read
\bes
S &=& \half \int dt\ (\dot q^2(t) - \om^2 q^2(t)) \nle 
\EE(t) &\equiv& -{\dlt S\/\dlt q(t)} = \ddot q(t) + \omega^2 q(t) = 0.
\eens
Introduce antifields $q^*(t)$ and canonical momenta $p(t) = \dd{q(t)}$
and  $p^*(t) = \dd{q^*(t)}$. The space spanned by  $q(t)$, $q^*(t)$, 
$p(t)$ and $p^*(t)$ is the extended phase space $\PP^*$.
The non-zero brackets are
\be
[p(t),q(t')] = [p^*(t),q^*(t')]_+ = \dlt(t-t').
\ee
The KT differential and the Hamiltonian in $\PP^*$  read
\bes
Q &=& \int dt\ (\ddot q(t) + \omega^2 q(t))p^*(t), 
\nle
H &=& -i\int dt\ (\dot q(t)p(t) + \dot q^*(t)p^*(t)).
\eens
$Q$ acts as $\dlt F = [Q,F]$, where
\bes
\dlt q(t) &=& 0, \nl
\dlt q^*(t) &=& \ddot q(t) + \omega^2 q(t), \nle
\dlt p(t) &=& -(\ddot p^*(t) + \omega^2 p^*(t)), \nl
\dlt p^*(t) &=& 0.
\eens

After Fourier transformation, $\PP^*$ is spanned by modes
$q_m$, $q^*_m$, $p_m$ and $p^*_m$, and the EL equations read
\be
\EE_m = -(m^2 - \omega^2)q_m = 0.
\ee
The non-zero Poisson brackets in $\PP^*$ are
\be
[p_m, q_n] = [p^*_m, q^*_n] = \dlt_{m+n}.
\ee
The KT differential and the Hamiltonian are
\bes
Q &=& \sum_{m=-\infty}^\infty (m^2 - \omega^2) q_m p^*_{-m}, 
\nlb{Qharm}
H &=& \sum_{m=-\infty}^\infty m (q_m p_{-m} + q^*_m p^*_{-m}).
\eens
$Q$ acts as $\dlt F = [Q,F]$, where
\bes
\dlt q_m &=& 0, \nl
\dlt q^*_m &=& (m^2 - \omega^2)q_m, 
\nlb{Qqp}
\dlt p_m &=& -(m^2 - \omega^2) p^*_m, \nl
\dlt p^*_m &=& 0.
\eens

The cohomology is computed as follows. 
Since the equations (\ref{Qqp}) decouple, we can consider each value
of $m$ separately. First assume that $m^2 \neq \omega^2$, i.e.
$m \neq \pm\omega$.
$q_m$ and $p^*_m$ are closed for all $m$, but $q^*_m$ and $p_m$
are not closed since $\dlt q^*_m \neq 0$, etc.
We can invert the second and third equations to read
\bes
q_m &=& {1\/m^2 - \omega^2} \dlt q^*_m, \nle
p^*_m &=& -{1\/m^2 - \omega^2} \dlt p_m.
\eens
Hence $q_m$ and $p^*_m$ lie in the image of $\dlt$,
and the cohomology vanishes completely: only $q_m$ and $p^*_m$ lie in
the kernel, but they also lie in the image.

Now turn to the case $m^2 = \omega^2$, say $m=\om$. Clearly,
$\dlt q_\om = \dlt p_\om = \dlt q^*_\om = \dlt p^*_\om = 0$, so
all four variables lie in the kernel but not in the image.
This clearly that the cohomology spaces are too big;
the classical cohomology spaces can be identified with 
$H_\cl^\bullet(\dlt) = C(q_{\pm\om}, p_{\pm\om}, q^*_{\pm\om},
p^*_{\pm\om})$. 
The zeroth cohomology space consists of such functions
with total antifield number zero, i.e. 
$H_\cl^0(\dlt) = C(q_{\pm\om}, p_{\pm\om}, (q^*_{\pm\om}p^*_{\pm\om}))$.

Now quantize by introducing a Fock vacuum $\ket 0$ satisfying
\be
q_{-m}\ket 0 = p_{-m}\ket 0 = q^*_{-m}\ket 0 = p^*_{-m}\ket 0 = 0,
\ee
for all $-m < 0$.
These conditions eliminate all modes with negative frequency $-\om$.
The quantum cohomology thus consists of Hilbert spaces built from the
modes with positive frequency $\om$,
$H_\qm^\bullet(Q) = C(q_{\om}, p_{\om}, q^*_{\om}, p^*_{\om})$.
In particular, the zeroth cohomology space consists of such functions
with total antifield number zero, i.e. 
$H_\qm^0(Q) = C(q_{\om}, p_{\om}, (q^*_{\om}p^*_{\om}))$.

This Hilbert space is also too big. It is spanned by states of the form
\be
\ket{n_q, n_p, n^*_q, n^*_p} 
= (q_\om)^{n_q} (p_\om)^{n_p} (q^*_\om)^{n^*_q} (p^*_\om)^{n^*_p} \ket 0.
\ee
The energy of this state is $(n_q + n_p + n^*_q +n^*_p)\om$, because
\be
H \ket{n_q, n_p, n^*_q, n^*_p} =
(n_q + n_p + n^*_q +n^*_p)\om\ \ket{n_q, n_p, n^*_q, n^*_p}.
\ee
The zeroth cohomology again consists of states with antifield number zero,
i.e. $n^*_q = n^*_p$.

One way to avoid the unwanted cohomology is to add a small perturbation,
so the Hessian
$K_{mn}(q) = \dlt^2 S/\dlt q_m \dlt q_n$ is non-singular and has an
inverse $M_{mn}(q)$. The $p$-part of (\ref{Qqp}) is replaced by
\bes
\dlt p_m &=& -\sum_n K_{mn}(q) p^*_n, \nle
\dlt p^*_m &=& 0,
\eens
i.e.
\bes
p^*_m &=& -\dlt( \sum_n M_{mn}(q) p_n), \nle
\dlt p^*_m &=& 0,
\eens
In the perturbed theory, $p^*_m$ is both closed and exact for all $m$, so
it vanishes in cohomology. Moreover, $p_m$ is never closed for any $m$.
After removing the small perturbation, the classical cohomology space can
thus be identified with
$H_\cl^\bullet(Q) = C(q_{\om}, q_{-\om}, q^*_{\om}, q^*_{-\om})$, and
the zeroth cohomology space is $H_\cl^0(Q) = C(q_{\om}, q_{-\om})$.
After quantization, the negative frequency modes $q_{-\om}$ and 
$q^*_{-\om}$ annihilate the vacuum, so the total quantum cohomology is
$H_\qm^\bullet(Q) = C(q_{\om}, q^*_{\om})$. In particular,
$H_\qm^0(Q) = C(q_{\om}) = \HH$ is the Hilbert space of the harmonic
oscillator. It has a basis consisting of $n$-quanta states,
$\ket n = (q_\om)^n \ket 0$ with the right energy, $H\ket n = n\om\ket n$.

One may note that the standard antifield treatment of the harmonic 
oscillator, without momenta, suffers from an analogous problem. 
Since $\dlt q_{\pm\om} = \dlt q^*_{\pm\om} = 0$, the $k$:th
cohomology group, rather than being zero, is spanned by functions of 
the form
$\sum_j {k\choose j}f_j(q_\om,q_{-\om})(q^*_{\om})^j(q^*_{\om})^{k-j}$.
The zeroth cohomogy space is thus the right physical phase space,
$H_\cl^0(\dlt) = C(q_\om,q_{-\om}) = C(\Sigma)$, but this is not a 
resolution because the higher cohomology groups do not vanish.

\section{Free scalar field: non-covariant quantization}

The action, Euler-Lagrange equations, and Hessian read
\be
S &=& \half \int \dNx (\dmu \phi(x)\d^\mu \phi(x) - \om^2 \phi^2(x)), \nl
\EE(x) &\equiv& -{\dlt S\/\dlt \phi(x)} 
= \dAlam \phi(x) + \omega^2 \phi(x) = 0, 
\nlb{Sscalar} 
K(x,x') &\equiv& -{\dlt^2 S\/\dlt \phi(x)\dlt \phi(x')} 
= \dAlam \delta(x-x') + \omega^2 \delta(x-x')
\eens
where $\dAlam = \dmu \d^\mu$.

Introduce antifields $\fs(x)$ and canonical momenta 
$\pi(x) = \dd{\phi(x)}$ and  $\ps(x) = \dd{\fs(x)}$.
The non-zero brackets are
\be
[\pi(x),\phi(x')] = [\ps(x),\fs(x')]_+ = \dlt(x-x').
\ee
The KT differential reads
\be
Q = \int \dNx (\dAlam \phi(x) + \omega^2 \phi(x))\ps(x). 
\ee
$Q$ acts as $\dlt F = [Q,F]$, where
\bes
\dlt \phi(x) &=& 0, \nl
\dlt \fs(x) &=& \dAlam \phi(x) + \omega^2 \phi(x), \nle
\dlt \pi(x) &=& -(\dAlam \ps(x) + \omega^2 \ps(x)), \nl
\dlt \ps(x) &=& 0.
\eens

Now we do a Fourier transformation. The extended phase space $\PP^*$
is spanned by modes $\phi(k)$, $\fs(k)$, $\pi(k)$ and $\ps(k)$, and
the EL equation becomes
\be
\EE(k) = -(k^2 - \omega^2)\phi(k) = 0.
\ee
The non-zero brackets are
\be
[\pi(k), \phi(k')] = [\ps(k), \fs(k')]_+ = \dlt(k-k').
\ee
The KT differential is
\be
Q = \int \dNk (k^2 - \omega^2) \phi(k) \ps(-k).
\ee
$Q$ acts as $\dlt F = [Q,F]$, where
\bes
\dlt \phi(k) &=& 0, \nl
\dlt \fs(k) &=& (k^2 - \omega^2)\phi(k), \nle
\dlt \pi(k) &=& -(k^2 - \omega^2) \ps(k), \nl
\dlt \ps(k) &=& 0.
\eens

To quantize the theory we must specify a Hamiltonian. Let it be
\bes
H &=& -i\int \dNx (\d_0 \phi(x)\pi(x) + \d_0 \fs(x)\ps(x)) \nle
  &=& \int \dNk k_0 (\phi(k) p(-k) + \fs(k) \ps(-k)).
\eens
Note that at this stage we break Poincar\'e invariance,
since the Hamiltonian treats the $x^0$ coordinate differently from
the other $x^\mu$.
Quantize by introducing a Fock vacuum $\ket 0$ satisfying
\be
\phi(k)\ket 0 = \pi(k)\ket 0 = \fs(k)\ket 0 = \ps(k)\ket 0 = 0,
\ee
for all $k$ such that $k_0 < 0$. 

The rest proceeds as for the harmonic oscillator. After adding a small
perturbation to make the Hessian invertible, $\pi(k)$ and $\ps(k)$ 
vanish in cohomology, as do the off-shell components of $\phi(k)$ 
and $\fs(k)$. The classical cohomology 
$H_\cl^\bullet(Q) = C(\phi(k; k^2 = \om^2), \fs(k; k^2 = \om^2))$ 
consists of functions of the on-shell components of $\phi$ and $\fs$, and 
$H_\cl^0(Q) = C(\phi(k; k^2 = \om^2))$ is the classical phase space.
The quantization step eliminates the components $\phi(k)$ with $k_0 < 0$,
which leaves us with the physical Hilbert space
$\HH = H_\qm^0(Q) = C(\phi(k; k^2 = \om^2\ \hbox{and}\ k_0 > 0))$.
A basis for $\HH$ consists of multi-quanta states
\be
\ket{k, k', ..., k^{(n)}} = \phi(k)\phi(k')...\phi(k^{(n)})\ket 0
\ee
with energy $H = k + k' + ... + k^{(n)}$.

\section{Free scalar field: covariant quantization}

Following the prescription in Section \ref{sec:covar}, we make the
replacement $\phi(x) \to \phi(x,t)$, where $t\in\RR$ is a parameter.
The EL equation (\ref{Sscalar}) becomes
\be
\EE(x,t) \equiv \dAlam \phi(x,t) + \omega^2 \phi(x,t) = 0.
\ee
To remove this condition in cohomology we introduce antifields $\fs(x,t)$.
But there is an extra condition
\be
\d_t\phi(x,t) \equiv {\d\phi(x,t)\/\d t} = 0.
\ee
We can implement this condition by introducing new antifields $\wf(x,t)$.
However, the identities $\d_t\EE(x,t) \equiv 0$ give rise to unwanted
cohomology. To kill this condition, we must introduce a second-order
antifield $\wfs(x,t)$.
The full KT differential $\dlt$ is now defined by
\bes
\dlt \phi(x,t) &=& 0, \nl
\dlt \fs(x,t) &=& \dAlam \phi(x,t) + \omega^2 \phi(x,t), \nle
\dlt \wf(x,t) &=& \d_t\phi(x,t), \nl
\dlt \wfs(x,t) &=& \d_t\fs(x,t) - (\dAlam \wf(x,t) + \omega^2 \wf(x,t)).
\eens

Introduce canonical momenta for all fields and antifields:
$\pi(x,t) = \dd{\phi(x,t)}$, $\ps(x,t) = \dd{\fs(x,t)}$,
$\wp(x,t) = \dd{\wf(x,t)}$, and $\wps(x,t) = \dd{\wfs(x,t)}$.
The KT differential can now be expressed as a bracket, $\dlt F = [Q,F]$,
where the KT operator is
\be
Q &=& \int \dNx \int dt\ \Big(
  (\dAlam \phi(x,t) + \omega^2 \phi(x,t))\ps(x,t)
  + \d_t\phi(x,t)\wp(x,t) 
\nl
&&+( \d_t\fs(x,t) - (\dAlam \wf(x,t) + \omega^2 \wf(x,t)))\wps(x,t) \Big).
\ees
Make a Fourier transform in $t$, e.g.
\be
\phi(x,t) = \intdm \phi(x,m)\e^{imt}.
\ee
which gives us Fourier-transformed fields
$\phi(x,m)$, $\fs(x,m)$, $\wf(x,m)$, $\wfs(x,m)$, on which the 
KT differential acts as
\bes
\dlt \phi(x,m) &=& 0, \nl
\dlt \fs(x,m) &=& \dAlam \phi(x,m) - \omega^2\phi(x,m), 
\nlb{dltm}
\dlt \wf(x,m) &=& m\phi(x,m), \nl
\dlt \wfs(x,m) &=& m\fs(x,m) - (\dAlam \wf(x,m) + \omega^2 \wf(x,m)).
\eens

The candidate Hamiltonian (\ref{Hamconstr}) acts on the fields as
\bes
[H_0, \phi(x,m)] = m\phi(x,m), &\qquad&
[H_0, \fs(x,m)] = m\fs(x,m), \nle
[H_0, \wf(x,m)] = m\wf(x,m), &\qquad&
[H_0, \wfs(x,m)] = m\wfs(x,m).
\eens
Alas, this Hamiltonian is identically zero on the physical phase space.
The third equation in (\ref{dltm}) can be rewritten as
$\phi(x,m) = m^{-1}\dlt\wf(x,m)$, which means that $\phi(x,m)$ is
KT exact and thus vanishes in cohomology unless $m=0$, and in that case
$H_0 = 0$. $H_0$ is thus a Hamiltonian constraint rather than 
a proper Hamiltonian.

To construct the physical Hamiltonian, we introduce the observer's
trajectory $ q^\mu(t) \in \RR^N$, and then expand all fields in a Taylor
series around this trajectory trajectory, i.e. we pass to jet data.
Hence e.g.,
\be
\phi(x,t) = \sum_{\mm} {1\/\mm!} \fm(t)(x-q(t))^\mm.
\label{STaylor}
\ee
The equation of motion and the time-independence condition translate
into
\bes
\sum_\mu \phi_{\mm+2\mu}(t) + \omega^2 \fm(t) &=& 0, \nle
D_t\fm(t) \equiv \ddt\fm(t) - \sum_\mu \dot q^\mu(t)\phi_{\mm+\mu} &=& 0.
\eens
We introduce anti-jets $\fsm(t)$, $\wfm(t)$ and $\wfsm(t)$ and the
KT differential $\dlt$ to implement these conditions:
\bes
\dlt \fm(t) &=& 0, \nl
\dlt \fsm(t) &=& \sum_\mu \phi_{\mm+2\mu}(t) + \omega^2 \fm(t), 
\nle
\dlt \wfm(t) &=& D_t\fm(t), \nl
\dlt \wfsm(t) &=& D_t\fs(t) 
 - (\sum_\mu \wf_{\mm+2\mu}(t) + \omega^2 \wfm(t)).
\eens
The classical cohomology group $H_\cl^0(\dlt)$ consists of linear 
combinations of jets satisfying
\be
\fm(t) = \e^{ik\cdot q(t)} (ik)^\mm
\label{fmt}
\ee
where $k^2 = \om^2$, $k\cdot q = k_\mu q^\mu$ and the power
$k^\mm$ is defined in analogy with (\ref{power}).
It is hardly surprising that the Taylor series (\ref{STaylor}) can be
summed, giving
\bes
\phi(x,t) &=&
\e^{ik\cdot q(t)} \sum_{\mm} {1\/\mm!} (ik)^\mm(x-q(t))^\mm \nl
&=& \e^{ik\cdot q(t)} \e^{ik\cdot(x-q(t))}
\label{phixt}\\
&=& \eikx.
\eens

The physical Hamiltonian $H$ is defined as in Equation (\ref{Hq}). The
classical phase space $H_\cl^0(\dlt)$ is thus the the space of plane waves
$\eikx$, cf (\ref{phixt}),  and trajectories $ q^\mu(t) = u^\mu t+a^\mu$.
The energy is given by
\bes
[H, \eikx] &=& k_\mu \dot q^\mu(t)\eikx = k_\mu u^\mu \eikx, \nle
[H,  q^\mu(t)] &=& i\dot q^\mu(t).
\eens
This is a covariant description of phase space, because the energy
$k_\mu u^\mu$ is Poincar\'e invariant.

We now quantize the theory before imposing the dyna\-mics. To this end, we
introduce the canonical momenta $\pim(t)$, $\psm(t)$, $\wpm(t)$,
$\wpsm(t)$ for the jets and antijets, and $ p_\mu(t)$ and $\psmu(t)$ for the
observer's trajectory and its antifield. The defining relations are
\bes
[\pim(t), \fn(t')] = \dlt^\mm_\nn \dlt(t-t'), &\quad&
[\psm(t), \fsn(t')] =  \dlt^\mm_\nn \dlt(t-t'), \nl
{[}\wpm(t), \wfn(t')] =  \dlt^\mm_\nn \dlt(t-t'), &\quad&
[\wpsm(t), \wfsn(t')] =  \dlt^\mm_\nn \dlt(t-t'), \nl
{[}p_\nu(t),  q^\mu(t')] = \dlt^\mu_\nu \dlt(t-t'), &\quad&
{[}\psmu(t), \qsnu(t')] = \dlt^\mu_\nu \dlt(t-t').
\label{LER2}
\ees
Since the jets also depend on the parameter $t$, we can define
their Fourier components, e.g. 
\be
\fm(t) = \intdm \fm(m)\e^{imt}, \qquad
 q^\mu(t) = \intdm  q^\mu(m)\e^{imt}.
\ee
The the Fock vacuum $\ket 0$ is defined to be annihilated by
the negative frequency modes of the jets and antijets, i.e.
\bes
&&\fm(-m)\ket 0 = \fsm(-m)\ket 0 = \wfm(-m)\ket 0 = \wfsm(-m)\ket 0 = 
\nl
&&\pim(-m)\ket 0 = \psm(-m)\ket 0 = \wpm(-m)\ket 0 = \wpsm(-m)\ket 0 =
\nl
&& q^\mu(-m)\ket 0 =  p_\mu(-m)\ket 0 = \qsmu(-m)\ket 0 = \psmu(-m)\ket 0 = 0.
\ees
for all $-m < 0$.
The quantum Hamiltonian is still defined by (\ref{Hq}),
where double dots indicate normal ordering with respect to frequency,
ensuring that $H\ket 0 = 0$.

The rest proceeds as in the end of Section \ref{sec:covar}.
We can consider the one-quantum state with momentum $k$ over the
true Fock vacuum, $\ket k = \exp(ik\cdot x)\ket 0$. This state is
not an energy eigenstate, because the Hamiltonian excites a quantum
of the observers trajectory: $H\ket k = k_\mu u^\mu\ket k$.
We may treat the observer's trajectory as a classical 
variable and introduce the macroscopic reference state 
$\ket{0;u,a}$, on which 
$ q^\mu(t)\ket{0;u,a} = (u^\mu t + a^\mu)\ket{0;u,a}$.
We can then consider a state
$\ket{k; u,a} = \exp(ik\cdot x)\ket{k; u,a}$ with one quantum over
the reference state. The Hamiltonian gives
$H \ket{k; u,a} = k_\mu u^\mu\ket{k;u,a}$.
In particular, if $u^\mu = (1,0,0,0)$, then
the eigenvalue of the Hamiltonian is $k_\mu u^\mu = k_0$, as
expected. Moreover, the lowest-energy condition (\ref{LER2}) ensures
that only quanta with positive energy will be excited; if 
$k_\mu u^\mu < 0$ then $\ket{k;u,a} = 0$.

\section{Reparametrization algebra}
\label{sec:repar}

In the covariant approach we introduced an auxiliary parameter $t$, which
is related to physical time through the geodesic equation (\ref{geo}). In
Minkowski space, the condition $\ddot q^\mu(t) = 0$ explicitly breaks
reparametrization invariance, and the reparametrization group does hence
not act in a well-defined manner on the cohomology. However, if we would
apply the formalism to a general-covariant theory such as general
relativity, the geodesic equation would not break reparametrization
invariance. The reason is that the metric, as one of the physical fields
$\fa(x)$, is replaced by a parametrized field $\fa(x,t)$. It is therefore
of interest to study the group of reparametrizations.

The infinitesimal generators are $L_f$, where $f = f(t)d/dt$ is a
vector field on the line. The reparametrization algebra 
acts on the parametrized fields as
\bes
{[}L_f, \fa(x,t)] &=& -f(t)\d_t\fa(x,t), \nl
{[}L_f, \fsa(x,t)] &=& -f(t)\d_t\fsa(x,t), \\
{[}L_f, q^\mu(t)] &=& -f(t)\dot q^\mu(t), 
\eens
etc. This translates into an action on the jet data:
\bes
[L_f, \fam(t)] = -f(t)\ddt\fam(t), &\quad&
{[}L_f, \fsam(t)] = -f(t)\ddt\fsam(t), \nl
{[}L_f, \pam(t)] = -f(t)\ddt\pam(t), &\quad&
{[}L_f, \psam(t)] = -f(t)\ddt\psam(t), \\
{[}L_f,  q^\mu(t)] = -f(t)\dot q^\mu(t), &\quad&
{[}L_f,  p_\mu(t)] = -f(t)\dot p_\mu(t),
\eens
etc.
A crucial observation is that $L_f$ is not a well defined operator on
the Hilbert space $\HH(\PP^*)$, because infinities arise when it acts
on the Fock vacuum. To remedy this, we normal order. However, $L_f$ is
still not a well-defined operator, because normal ordering formally
gives rise to an {\em infinite} central extension.
We must therefore regularize the generators further. Fortunately, a
natural regularization is available in jet space: simply truncate
the Taylor expansion at some fixed, finite order $p$. In other words,
we pass from the space of infinite jets to the space of $p$-jets
(or rather trajectories in $p$-jet space). This regularization respects
all relevant symmetries, such a Poincar\'e, diffeomorphism or gauge
symmetry, and it is in fact the unique regularization with this 
property.

A normal-ordered and regularized generator of the reparametrization
algebra acting only on the proper fields is thus
\be
L_f = -\int dt\ f(t) ( \no{\dot q^\mu(t) p_\mu(t)}
+ \sum_{|\mm|\leq p} \no{\ddt\fam(t)\pam(t)} )
\label{Lf}
\ee
Such operators generate a Virasoro algebra,
\be
{[}L_f,L_g] = L_{[f,g]} 
 + {c\/24\pi i}\int dt (\ddot f(t) \dot g(t) - \dot f(t) g(t)), 
\ee
with central charge
\be
c = 2 + 2n\Np{},
\label{c}
\ee
where $n$ is the number of components $\fa(x,t)$ and $N$ is the 
dimension of spacetime.
The first term is the contribution from the trajectory $q^\mu(t)$ and
its momentum $p_\mu(t)$, and the second term comes from the $p$-jets
$\fam(t)$; one verifies that a multi-index $\mm$ can take on
$\Np{}$ different values with $|\mm|\leq p$ in $N$ dimensions.

A regularization must be removed at the end of the process, i.e. we must
must take the limit $p\to\infty$ in (\ref{Lf}). A necessary condition for
taking this limit is that the central charge converges to a finite value
in this limit. Taken at face value, the prospects for succeeding appear
bleak. When $p$ is large, $c = \Np{} \approx p^N/N!$, so
the central charge diverges.
However, there is a way out of this problem. A bosonic $p$-jet 
contributes $2\Np{}$ to $c$, but a fermionic one makes the same
contribution with opposite sign. In a theory with both bosonic and
fermionic jets of different order, the leading divergences can cancel,
leaving us with a finite $p\to\infty$ limit.

Consider $n_i$ $(p-i)$-jets, where the sign of $n_i$ decides the
Grassmann parity; $n_i>0$ for bosons and $n_i<0$ for fermions.
The reparametrization algebra acts as a direct sum on the space of
collections of jets, with different values of the jet order $p$. Take
the sum of $r+1$ terms like those in (\ref{c}), with $p$ replaced by
$p$, $p-1$, ..., $p-r$, respectively, and with $n$ replaced by
$n_i$ in the $p-i$ term. The total central charge is
\be
c_{TOT} = n_0\Np{} + n_1\Np{-1} + ... + n_r\Np{-r}.
\ee
Using the recurrence formula,
\be
{n\choose i} = {n\choose i-1} + {n-1\choose i-1},
\ee
it is straightforward to show that
\be
c_{TOT} = {N-r \choose N} n_0,
\ee
provided that
\be
n_i = n_0 (-)^i {r\choose i}.
\label{conds}
\ee

Such a sum of contributions arises naturally from the KT complex, because
the antifields are only defined up to an order smaller than $p$;
if $\fam(t)$ is defined for $|\mm|\leq p$, $\fsam(t)$ is only 
defined for $|\mm|\leq p-2$, because the EL equations are second order.
The EL jets $\Eam(t)$ with $|\mm|\geq p-1$ involves the field jets
$\fam(t)$ with $|\mm|\geq p+1$, which were truncated by the 
regularization.
In the full reparametrization algebra, the generator (\ref{Lf}) is
replaced by
\bes
L_f &=& -\int dt\ f(t) \Big( 
 \no{\dot q^\mu(t) p_\mu(t)} + \no{\dot\qsmu(t)\psmu(t)} \nl
&&+ \sum_{|\mm|\leq p} \no{\ddt\fam(t)\pam(t)} 
 + \sum_{|\mm|\leq p-2} \no{\ddt\fsam(t)\psam(t)} 
\label{Lf2}\\
&&+ \sum_{|\mm|\leq p-1} \no{\ddt\wfam(t)\wpam(t)} 
+ \sum_{|\mm|\leq p-3} \no{\ddt\wfsam(t)\wpsam(t)} \Big),
\eens
assuming that the EL equations are second order.

Let us now consider the solutions to (\ref{conds}) for the numbers
$n_i$, which can be interpreted as the number of fields and anti-fields.
First assume
that the field $\fam(t)$ is fermionic with $n_F$ components, which gives
$n_0=-n_F$. We may assume, by the spin-statistics theorem, that the 
EL equations are first order, so the bosonic antifields $\fsam(t)$ 
contribute $n_F$ to $n_1$. The barred antifields $\wfam(t)$ are also
defined up to order $p-1$, and so give $n_1=n_F$, and the barred 
second-order antifields $\wfsam(t)$ give $n_2=-n_F$. 
For bosons the situation is analogous, with two exceptions: all signs are
reversed, and the EL equations are assumed to be second order. Hence
$\fsam(t)$ yields $n_2=-n_B$, and the barred antifields are one order
higher.

The situation is summarized in the following tables, where the upper half
is valid if the original field is fermionic and the lower half if it is
bosonic:
\bes
\barr{|c|c|c|l|}                                                       
\hline
i & \hbox{Jet} & \hbox{Order} & n_i \\
\hline
0 & \fam(t) & p & -n_F \\
1 & \wfam(t) & p-1 & n_F \\
1 & \fsam(t) & p-1 & n_F \\
2 & \wfsam(t) & p-2 & -n_F \\
\hline
\hline
0 & \fam(t) & p & n_B \\
1 & \wfam(t) & p-1 & -n_B \\
1 & \fsam(t) & p-2 & -n_B \\
2 & \wfsam(t) & p-3 & n_B \\
\hline
\earr
\label{tab}
\ees
If we add all contributions of the same order, we see that relation 
in (\ref{conds}) can only be satisfied provided that
\bes
p: &\quad& -n_F + n_B = n_0 \nl
p-1: && 2n_F - n_B = -rn_0, \nl
p-2: && -n_B - n_F = {r\choose2}n_0, 
\label{rcond}\\
p-3: && n_B = -{r\choose3}n_0, \nl
p-4: && 0 = {r\choose4}n_0, ..
\eens
The last equation holds only if $r\leq3$ (or trivially if $n_0=0$). 
On the other hand, if we demand that there is at least one bosonic
field, the $p-3$ equation yields $r\geq3$. Thus
we are unambigiously guided to consider
$r=3$ (and thus $N=3$). The specialization of (\ref{rcond}) to three
dimensions reads
\bes
p: &\quad& -n_F + n_B = n_0 \nl
p-1: && 2n_F - n_B = -3n_0, \nle
p-2: && -n_B - n_F = 3n_0, \nl
p-3: && n_B = -n_0,
\eens
Clearly, the unique solution to these equations is
\be
n_F = -2n_0, \qquad n_B = -n_0.
\label{xsol}
\ee
The negative sign implies that the sign of the central charge is 
negative. Thus, in three dimensions $c_{TOT} = -n_0$ independent of
the truncation order $p$, provided that the bosonic fields have
$n_0$ components and the fermionic fields have twice as many components.
Moreover, this is the only case where the total central charge has
a finite, non-zero limit when the regulator is removed, i.e.
the limit $p\to\infty$ can be taken.

The counting changes in the presence of gauge symmetries. It was 
observed in \cite{Lar02} that if all gauge symmetries are irreducible,
then the $p\to\infty$ limit exists only in four dimensions. This can
be viewed as a non-trivial prediction for the spacetime dimension.

\section{Conclusion}

The key observation in this paper is that we may obtain the physical
phase space by reduction from the phase space of histories $\PP$. The
dyna\-mics, i.e. the Euler-Lagrange equations, play the role of 
first-class constraints. This allows us to apply standard methods from the
theory of constrained Hamiltonian systems, e.g. Dirac brackets and
cohomological methods. To obtain a truly covariant formulation, we
expand all fields in a Taylor series around the observer's trajectory,
which acquires the status of a physical field, whose dyna\-mics is
governed by the geodesic equation. 

These methods were then applied to the harmonic oscillator and to the
free scalar field. The expected results were recovered if the vacuum is a
macroscopic reference state where the observer's velocity is a classical
variable. However, like all other physical objects, the observer's
trajectory is fundamentally quantum, and must be in principle be
treated as a quantum object. Observation is a special case of 
quantum interaction.

The reparametrization algebra can only act in a well-defined manner
provided that spacetime has three dimensions and there are twice as
many fermions as bosons. In the presence of gauge symmetries, the 
counting changes so that spacetime must have four dimensions, apparently
in good agreement with experiments.

There are several directions in which the present work can be extended.
The present construction is exact but not explicit; extracting numbers
will not be simpler than with standard methods. To extract numbers, we
need to formulate perturbation theory and understand renormalization
from this new viewpoint. Another generalization is to consider theories
with gauge symmetries, such as Yang-Mills theory and general relativity.
Essential calculations were done already in \cite{Lar02}, although
with some errors, particularly in nomenclature and physical
interpretation. 

A most striking new feature is the appearance of new anomalies, which
can not be seen in field theory because they are functionals of the
observer's trajectory. E.g., consider the algebra of maps from $\RR^N$
to a finite-dimensional Lie algebra $\oj$ with structure constants
$f^{ab}{}_c$ and Killing metric $\dlt^{ab}$. Let $X=X_a(x)J^a$ be a
generator of this current algebra, and $[X,Y]_c = f^{ab}{}_c X_aY_b$.
It will acquire an extension of the form
\be
{[}\J_X, \J_Y] = \J_{[X,Y]} - {k\/2\pi i}\dlt^{ab}
 \int dt\ \dot q^\rho(t)\d_\rho X_a(q(t))Y_b(q(t)),
\ee
which clearly generalizes affine algebras to several dimensions.
The constant $k$ is the value of the second Casimir operator. 
It vanishes in the special case of the adjoint representation of $u(1)$,
so quantization of the free Maxwell field will go through as usual,
but interacting gauge fields will necessarily acquire such new 
quantum corrections.
This clearly indicates that making the observer's trajectory physical
is not only an option, but absolutely necessary; a missed anomaly is
a serious oversight. This will be the subject of an upcoming publication.

\end{document}